\documentclass[
 aip,
 amsmath,amssymb,
 reprint,%
]{revtex4-1}

\usepackage{gensymb}
\usepackage{here}
\usepackage{graphicx}
\usepackage{dcolumn}
\usepackage{bm}
\usepackage{cases}
\usepackage{url}

\usepackage{xcolor}

\usepackage[T1]{fontenc}
\usepackage{newtxtext, newtxmath}
\usepackage{isomath}
\DeclareMathAlphabet{\mathbfit}{OML}{nxlmi}{bx}{it}
\let\lvec\undefined

\newcommand{\Alfven}{Alfv\'en }

\newcommand{\Wce}{\Omega_{e}}

\newcommand{\tbn}{\theta_{\rm Bn}}
\newcommand{\VAe}{V_{\rm A,e}}

\newcommand{\MA}{M_{\rm A}}

\newcommand{\MAHTF}{M_{\rm A}^{\rm HTF}}
\newcommand{\lvec}{\mathbfit{l}}
\newcommand{\mvec}{\mathbfit{m}}
\newcommand{\nvec}{\mathbfit{n}}
\renewcommand{\bm}[1]{\mathbfit{#1}}

\graphicspath{{figure/}}

\begin{document}

\preprint{AIP/123-QED}

\title[High-frequency Whistler Waves at Bow Shock]{Statistical Analysis of High-frequency Whistler Waves at Earth's Bow Shock: Further Support for Stochastic Shock Drift Acceleration}

\author{Takanobu Amano}
\email{amano@eps.s.u-tokyo.ac.jp}
\author{Miki Masuda}
\affiliation{
Department of Earth and Planetary Science, The University of Tokyo, Tokyo 113-0033, Japan.
}
\author{Mitsuo Oka}
\affiliation{
Space Sciences Laboratory, University of California, Berkeley, California 94720, USA.
}
\author{Naritoshi Kitamura}
\affiliation{
Institute for Space-Earth Environmental Research, Nagoya University, Nagoya 464-0601, Japan.
}
\author{Olivier Le Contel}
\affiliation{
Laboratoire de Physique des Plasmas, CNRS/Sorbonne Universit\'e/Universit\'e Paris-Saclay/Observatoire de Paris/Ecole Polytechnique Institut Polytechnique de Paris, Paris, 75005 France.
}
\author{Daniel J. Gershman}
\affiliation{
NASA Goddard Space Flight Center, Greenbelt, Maryland 20771, USA
}

\date{\today}

\begin{abstract}
We statistically investigate high-frequency whistler waves (with frequencies higher than $\sim 10$ \% of the local electron cyclotron frequency) at Earth's bow shock using Magnetospheric Multi-Scale (MMS) spacecraft observations. We focus specifically on the wave power within the shock transition layer, where we expect electron acceleration via stochastic shock drift acceleration (SSDA) to occur associated with efficient pitch-angle scattering by whistler waves. We find that the wave power is positively correlated with both the \Alfven Mach number in the normal incidence frame $\MA$ and in the de Hoffmann-Teller frame $\MA/\cos\tbn$. The empirical relation with $\MA/\cos\tbn$ is compared with the theory of SSDA that predicts a threshold wave power proportional to $(\MA/\cos \tbn)^{-2}$. The result suggests that the wave power exceeds the theoretical threshold for $\MA / \cos \tbn \gtrsim 30-60$, beyond which efficient electron acceleration is expected. This aligns very well with previous statistical analysis of electron acceleration at Earth's bow shock (M. Oka, Geophys.~Res.~Lett., 33, 5, 2006). Therefore, we consider that this study provides further support for SSDA as the mechanism of electron acceleration at Earth's bow shock. At higher-Mach-number astrophysical shocks, SSDA will be able to inject electrons into the diffusive shock acceleration process for subsequent acceleration to cosmic-ray energies.
\end{abstract}

\maketitle

\section{Introduction}
\label{sec:intro}
Collisionless shock waves have been studied extensively over the decades. Contrary to those in ordinary collisional media, the dissipation required at the collisionless shock must be provided by collective plasma processes, such as wave-particle interactions. It is, therefore, not surprising that various kinds of plasma waves are observed in the shock transition layer (STL), where the major dissipation takes place \citep{Wilson2014b}. How the waves are generated, interact with particles, and eventually dissipate is a central subject of collisionless shock research.

Another motivation to study collisionless shocks is the acceleration of energetic charged particles. The standard paradigm of the origin of cosmic rays assumes that they are accelerated by supernova remnant (SNR) shocks \citep{Drury1983a,Blandford1987a}. Astrophysical multi-wavelength observations have revealed that efficient acceleration of electrons to ultra-relativistic energies is a common feature of young SNR shocks \citep{Reynolds2008a}. Although the standard diffusive shock acceleration (DSA) theory is capable of explaining the acceleration of relativistic electrons, it is known that DSA is inefficient for low energy suprathermal electrons \citep{Amano2022a}. The so-called electron injection problem states that there must be a pre-acceleration mechanism such that a fraction of electrons in the thermal pool are energized and injected into the DSA process for subsequent acceleration to ultra-relativistic energies. It should be noted that efficient acceleration of relativistic electrons is rarely seen by in-situ spacecraft observations of shocks in the heliosphere \citep{Gosling1989a,Shimada1998a,Lario2003a,Dresing2016a}, indicating that the efficiency of electron injection is likely dependent on the shock parameters.

Given the observed rich variety of plasma waves seen near the shock, it is natural to anticipate that the pre-acceleration mechanism is somehow related to the wave activity \citep{Cargill1988a,Papadopoulos1988a,Galeev1995a,Dieckmann2000a,McClements1997a,McClements2001a,Hoshino2002a,Amano2007a,Amano2009b,Riquelme2011a}. In fact, we have recently proposed a theory of electron injection at an oblique shock that assumes intense wave activity within STL \citep{Katou2019a,Amano2020a,Amano2022b}. It is called stochastic shock drift acceleration (SSDA), which is essentially shock drift acceleration (SDA) with an added effect of efficient pitch-angle scattering via wave-particle interaction. We have previously demonstrated the validity of the theory using observations of Earth's bow shock made by Magnetospheric Multi-Scale (MMS) spacecraft for a single bow shock crossing event \citep{Amano2020a}. Furthermore, several more examples have recently been analyzed to confirm the consistency with the theory \citep{Lindberg2023a}. Fully kinetic simulations have also found electron acceleration consistent with SSDA \citep{Matsumoto2017a,Kobzar2021a,Ha2021a}.

Although so far successful, there is yet a crucial element missing in the theory. It predicts a threshold wave power above which the electron acceleration becomes efficient. However, it does not predict the wave power itself. It is indeed quite challenging to estimate the wave amplitude in the violently evolving dynamical STL from an analytical approach. On the other hand, its dependence on the shock parameters has to be understood for applications to a wider range of shocks in the heliosphere and beyond, in particular, where detailed high-time-resolution measurements are not available.

The primary motivation of this paper is to investigate the statistical property of electromagnetic wave power using MMS spacecraft observations. We specifically focus on the wave power in the whistler wave frequency range, which we expect to be the most relevant for the scattering of suprathermal electrons observed at Earth's bow shock. We study correlations between the wave power and the shock parameters, particularly the \Alfven Mach number and the magnetic obliquity angle. We argue that the obtained empirical relation supplemented by the theory of SSDA \citep{Katou2019a,Amano2020a,Amano2022b} well explains the existing statistical analysis of electron acceleration by \citet{Oka2006a}. This provides further support for SSDA as the mechanism of electron acceleration at Earth's bow shock and, ultimately, the solution to the electron injection problem.

The remainder of this paper is organized as follows. Section \ref{sec:data_and_method} introduces the data and method of analysis used in this study. Readers not interested in the detailed description of data analysis may directly jump into Section \ref{sec:discussion}, in which the result of statistical analysis is presented and discussed. Finally, the summary and conclusion are given in Section \ref{sec:conclusion}.

\section{Data and Method}
\label{sec:data_and_method}

\subsection{Dataset}
\label{sec:dataset}
We utilized Earth's bow shock crossing data obtained by MMS spacecraft \citep{Burch2016a}. The wave power of the magnetic fluctuations was calculated using burst-mode data from the search-coil magnetometer (SCM) \citep{LeContel2016a}, while the fluxgate magnetometer (FGM) \citep{Russell2016a} was used for the DC magnetic field. In addition, for the determination of the shock parameters, the plasma moments obtained by the fast plasma investigation (FPI) \citep{Pollock2016a} and the magnetic field obtained by FGM in fast survey mode were used because the burst mode data were often unavailable in the upstream and downstream regions of the shock. Note that we took $4.5 \, {\rm s}$ average of the original 16 Hz sampling magnetic field measurement by FGM in fast survey mode to match the time resolution of the FPI moment quantities. Since the MMS measurement of the density and pressure in the solar wind may not always be accurate in the upstream of the shock, we also used the OMNI data from NASA's OMNIWeb to check the solar wind parameters.

We selected 121 candidate shock-crossing events that occurred between the beginning of 2016 and the end of 2017, for which the burst mode data is available for the entire shock transition. We note that the event selection was somewhat biased toward the quasi-perpendicular geometry because we chose the events that look better suited for the shock parameter estimate. Quasi-parallel shocks are usually much more turbulent and often accompanied by large-amplitude foreshock disturbances, which makes it difficult to apply the automated shock parameter estimate method as described below. Note that even if we included those events in the candidate list, they would be eventually discarded during the event selection procedure, and the final dataset used for the statistical analysis should have been nearly the same.

\subsection{Shock Parameters}
\subsubsection{Shock Geometry}
\label{sec:geometry}
\begin{figure}[htbp]
  \centering
  \includegraphics[width=0.50\textwidth]{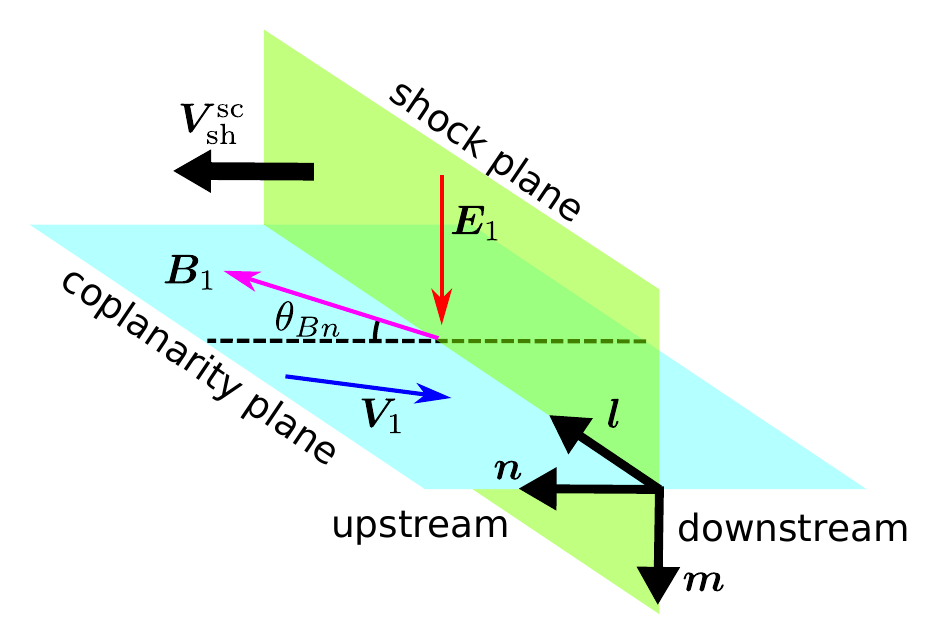}
  \caption{Definition of the $LMN$ coordinate system used in this study. Both the magnetic field vector and the velocity vector are contained within the $\nvec-\lvec$ plane, also known as the coplanarity plane. The electric field is perpendicular to the coplanarity plane and is parallel to $\mvec$. In the spacecraft frame, the shock wave propagates with a speed of $V_{\rm sh}^{\rm sc}$ in the $\nvec$ direction.}
  \label{fig:shock_geometry}
\end{figure}

First, it is important to estimate the shock geometry because our primary purpose is to investigate the dependence of wave power on the shock parameters in a statistical sense. In other words, the shock normal vector and the two orthogonal vectors in the plane of the shock surface must be defined appropriately. As shown in Fig.~\ref{fig:shock_geometry}, we introduced a local $LMN$ coordinate system defined as follows: the $N$-axis is the shock normal direction and pointing away from the Earth (i.e., positive toward upstream), the $L$-axis is parallel to the transverse magnetic field, and the $M$-axis is defined such that the resulting coordinate system becomes right-handed, or $\nvec = \lvec \times \mvec$ where $\lvec$, $\mvec$, $\nvec$ are unit vectors corresponding to each axis. As we shall see below, for each event, we performed a frame transformation to the normal incidence frame (NIF) to visually check the quality of the chosen coordinate system. In this frame, the upstream plasma flow is anti-parallel to $\nvec$, the magnetic field is contained in the $\nvec-\lvec$ plane, and the motional electric field $\bm{E} = - \bm{V} \times \bm{B}$ is parallel to $\mvec$. Because of the coplanarity theorem for an MHD shock, the plasma flow velocity vector and the magnetic field vector in both the far upstream and downstream regions must be in the $\nvec-\lvec$ plane, which is known as the coplanarity plane.

There have been a number of methods proposed to estimate the geometry of the shock \citep{Schwartz1998a}, but none of them are perfect. For the statistical study, a simple yet robust method that is not too sensitive to the shock parameters (such as obliquity) and intrinsic errors of the measurements is preferable. We consider only single spacecraft techniques because systematic errors may arise with multi-spacecraft methods \citep{Russell1983a,Russell1983b} when applied automatically to events with different spacecraft separations. We found that the conventional minimum variance analysis (MVA) and the magnetic coplanarity (MC) are sensitive to the chosen interval of analysis and often give estimates inconsistent with the other methods, in agreement with earlier reports \citep{Horbury2001a,Shen2007a}. The method of \citet{Vinas1986a} (VS) appears to be more comprehensive and has been used in the literature. It uses a subset of Rankine-Hugoniot relations; the conservation of mass flux, transverse momentum, magnetic flux, and the normal magnetic field component. However, since we are aware that the ion and electron densities obtained by FPI sometimes do not match each other, we are concerned about systematic errors in the density measurement. We have found that simpler methods based on the coplanarity but use both the velocity and magnetic field \citep{Abraham-Shrauner1972a,Abraham-Shrauner1976a} often give reasonable estimates consistent with the VS method. More specifically, the normal vector $\nvec$ may be given by
\begin{align}
    \nvec &= \pm
    \frac
    {\left(\bm{\tilde{B}} \times \Delta \bm{V}\right) \times \Delta \bm{B}}
    {\left| \left(\bm{\tilde{B}} \times \Delta \bm{V}\right) \times \Delta \bm{B} \right|},
    \label{eq:normalvector}
\end{align}
where $\Delta \bm{V} = \bm{V}_2 - \bm{V}_1$, $\Delta \bm{B} = \bm{B}_2 - \bm{B}_1$, and $\bm{\tilde{B}}$ may be either $\bm{B}_1$ or $\bm{B}_2$. In the following, the quantities with the subscript $1$ and $2$ always refer to those in the upstream and downstream, respectively. Note that we chose the sign such that $\nvec$ always points radially outward in the Geocentric Solar Ecliptic (GSE) coordinate system, which corresponds to the upstream direction at Earth's bow shock. We will refer to the methods as AY1 and AY2 after \citet{Abraham-Shrauner1976a}, who called the methods the mixed data 1 (MD1) and 2 (MD2). Note that the coplanarity theorem results from the same subset of Rankine-Hugoniot relations used in the VS method. It is thus not surprising that the AY method gives similar results to the VS method. The advantage of the AY method is that it does not require the density measurement and can be applied to the observation in an arbitrary reference frame because only the velocity difference is involved in Eq.~\ref{eq:normalvector}.

Once the normal vector $\nvec$ is determined, the two orthogonal vectors $\lvec$ and $\mvec$ are readily determined by
\begin{align}
    \lvec &=
    \frac
    {\bm{\tilde{B}} - \left( \bm{\tilde{B}} \cdot \nvec \right) \nvec}
    {\left| \bm{\tilde{B}} - \left( \bm{\tilde{B}} \cdot \nvec \right) \nvec \right|}, \\
    \mvec &= \frac{\nvec \times \lvec}{\left| \nvec \times \lvec \right|}.
\end{align}
Therefore, depending on the choice of $\bm{\tilde{B}}$ (either $\bm{B}_1$ or $\bm{B}_2$), two complete right-handed coordinate systems are obtained from the velocity and magnetic field measurements of given upstream and downstream intervals.

In practice, we chose the upstream and downstream intervals such that the two methods (AY1 and AY2) give consistent estimates of the shock normal vector as much as possible. More specifically, the best pair of intervals was determined as follows. First, we fixed the length of upstream and downstream time interval to $13.5 \, {\rm s}$ (or 3 data points with FPI time resolution of the fast survey mode $4.5 \, {\rm s}$). The candidate upstream and downstream time intervals were selected by continuously sliding the time window both forward and backward in time up to $90 \, {\rm s}$, starting from manually pre-defined upstream and downstream edges of the shock transition. For each candidate pair of intervals, we applied the AY methods to all the possible combinations of upstream and downstream pairs of data points (9 combinations for 3 upstream and downstream data points, respectively). We determined the best pair of intervals such that the mean difference between the two AY methods is minimized over the interval. Once the interval was selected, we obtained the final estimate by taking the mean. Similarly, the final error was defined by the standard deviation.

\subsubsection{Shock Propagation Velocity}
\label{sec:velocity}
One of the most natural frames to define the shock propagation velocity is NIF. In this case, the shock velocity is defined by the shock propagation velocity normal to the shock surface in the upstream plasma rest frame $\bm{V}_{\rm sh} = V_{\rm sh} \nvec$ (notice that $V_{\rm sh} > 0$ by definition). Since the position of Earth's bow shock is always fluctuating in response to the solar wind, we need a reasonable estimate of the instantaneous shock velocity in the spacecraft frame $V_{\rm sh}^{\rm sc}$ to obtain $V_{\rm sh} = - \bm{V}_1 \cdot \bm{n} - V_{\rm sh}^{\rm sc}$ (see Fig.\ref{fig:shock_geometry}).

Again, we prefer a method that requires only the velocity and magnetic field measurements. For this purpose, we may use magnetic flux conservation. With the $LMN$ coordinate system appropriately defined, the conserved quantity is given by $E'_{m} = - \left(V'_{n} B_{l} - V'_{l} B_{n} \right)$ where the primed quantities are those defined in the shock rest frame (which may not necessarily be NIF as $V'_{l} \neq 0$). The spacecraft frame shock velocity is thus readily obtained by
\begin{align}
  \bm{V}_{\rm sh}^{\rm sc} = \frac{ \left[ V_{n} B_{l} - V_{l} B_{n} \right] }{ \left[ B_{l} \right] } \nvec,
  \label{eq:shockvelocity}
\end{align}
where $\left[ A \right] \equiv A_2 - A_1$ denotes the difference of a physical quantity $A$ between the upstream and downstream.

It is important to mention that the method, in the ideal case, is equivalent to the one proposed by \citet{Smith1988a} (SB) that is based on the same principle. More specifically, it is easy to show the equivalence when the following condition is met: $\bm{B}_2 - \bm{B}_1 = (B_{l,2} - B_{l,1}) \lvec$. In other words, the coplanarity ($B_{n,2} = B_{n,1}$ and $B_{m,2} = B_{m,1} = 0$) needs to be satisfied for the SB and the present method give the same result. This indicates that, although the SB method can blindly be applied to any event, it will give an erroneous result when the coplanarity is not well satisfied. For the statistical analysis presented in this paper, we manually checked the results and eliminated such events from the list (see Section \ref{sec:example}).

The shock velocity Eq.~\ref{eq:shockvelocity} can be evaluated for all the combinations in the best intervals both with either AY1 or AY2 (to determine the shock normal vector $\nvec$). With this distribution, we estimated $\bm{V}_{\rm sh}^{\rm sc}$ by the mean and its error by the standard deviation.

\subsubsection{Shock Parameters}
\label{sec:other}
One of the most important parameters known to regulate the shock dynamics is the magnetic field obliquity angle $\tbn$ (see Fig.~\ref{fig:shock_geometry}). With the estimated shock normal vector and the mean magnetic field in the upstream interval, we calculated $\cos \tbn = (\nvec \cdot \bm{B}_1) / |\bm{B}_1|$. Note that we consider $\cos \tbn$ (rather than $\tbn$ itself) as the primary quantity, including its sign. This is because (1) it appears directly in the following analysis, (2) otherwise, the definition of the error near $\cos \tbn = 0$ (or $\tbn = 90\degree$) can be somewhat misleading.

Two more physically important parameters characterizing an MHD shock are the \Alfven Mach number $\MA$ and the plasma beta $\beta$. We define the \Alfven Mach number in NIF, which is thus given by $M_{\rm A} = V_{\rm sh} / V_{A}$, where $V_{A} = 21.8 \, B_0 / \sqrt{n_0}$ is the \Alfven speed in units of ${\rm km/s}$ defined with the upstream magnetic field strength $B_0 = |\bm{B}_1|$ (in units of ${\rm nT}$) and the number density $n_0$ (in units of ${\rm cm^{-3}}$). As already mentioned, there may be a systematic error in the ion and electron density measurement by FPI in the solar wind. Therefore, the \Alfven Mach number slightly changes depending on which density we adopt for the definition of the \Alfven velocity. In the present paper, we simply use the \Alfven Mach number defined with the ion density, but we have confirmed that the results do not change significantly even if we use the electron density instead.

Accurate estimation of the upstream plasma beta is also an issue. It has to be mentioned that the automatic time interval selection procedure described earlier does not take into account possible contamination of the solar wind plasma quantities by the presence of the shock. Although the impact of the shock on the upstream density does not seem to be substantial, the upstream temperature may be affected by the shock-reflected particles as well as the pre-heating associated with the precursor wave activity. Furthermore, the angular resolution of the FPI instrument may not be sufficient to fully resolve the cold solar wind ion distribution. We have compared the upstream ion and electron plasma betas $\beta_i$, $\beta_e$ obtained by MMS with the OMNI data $\beta_{\rm omni}$ averaged over 1 hour and found a statistical correlation but with significant variance. In particular, the ion plasma beta $\beta_i$ tends to be larger than $\beta_{\rm omni} / 2$ (assuming that the ion and electron temperatures are the same $T_i/T_e = 1$), which is consistent with the above-mentioned artifacts. The electron plasma beta $\beta_e$, on the other hand, appears to be less contaminated. Therefore, we will use the electron beta $\beta_e$ obtained by MMS and the one-hour-averaged total plasma beta $\beta_{\rm omni}$ obtained by OMNI as a proxy in the following discussion. Nevertheless, we should keep in mind that any dependence on the plasma betas has to be interpreted with caution.

Whenever needed, the errors in $\cos \tbn$, $\MA$, $\beta_e$, $\beta_{\rm omni}$, and other parameters defined by combinations of these quantities were always evaluated by properly taking into account error propagation from those of primary quantities. We note that the errors were typically dominated by those of $\cos \tbn$ and $V_{\rm sh}^{\rm sc}$.

\subsubsection{Example}
\label{sec:example}
We can always obtain the shock geometry and the shock parameters with the method described above, which may, however, not always be physically reasonable. To confirm the consistency with the theoretical properties of a fast-mode MHD shock, we checked whether the following conditions are satisfied across the shock (but not necessarily within STL): (1) $B_{n} = {\rm const.}$ and $B_{m} = 0$, (2) $E_{m} = {\rm const.}$ and $E_{l} = E_{n} = 0$, where the electric field is defined in NIF (the magnetic field is invariant under Galilean transformation). The above condition indicates that the plasma flow velocity in NIF should satisfy the following property: (3) $V_{n}$ is the primary component of deceleration across the shock, (4) there may be a small but finite change in the transverse velocity in $V_{l}$, and (5) $V_{m}$ is constant. Because of the intrinsic fluctuations, it is not easy to check these conditions automatically. We therefore made plots of these quantities converted into NIF and inspected them one by one.

\begin{figure}[htbp]
  \centering
  \includegraphics[width=0.50\textwidth]{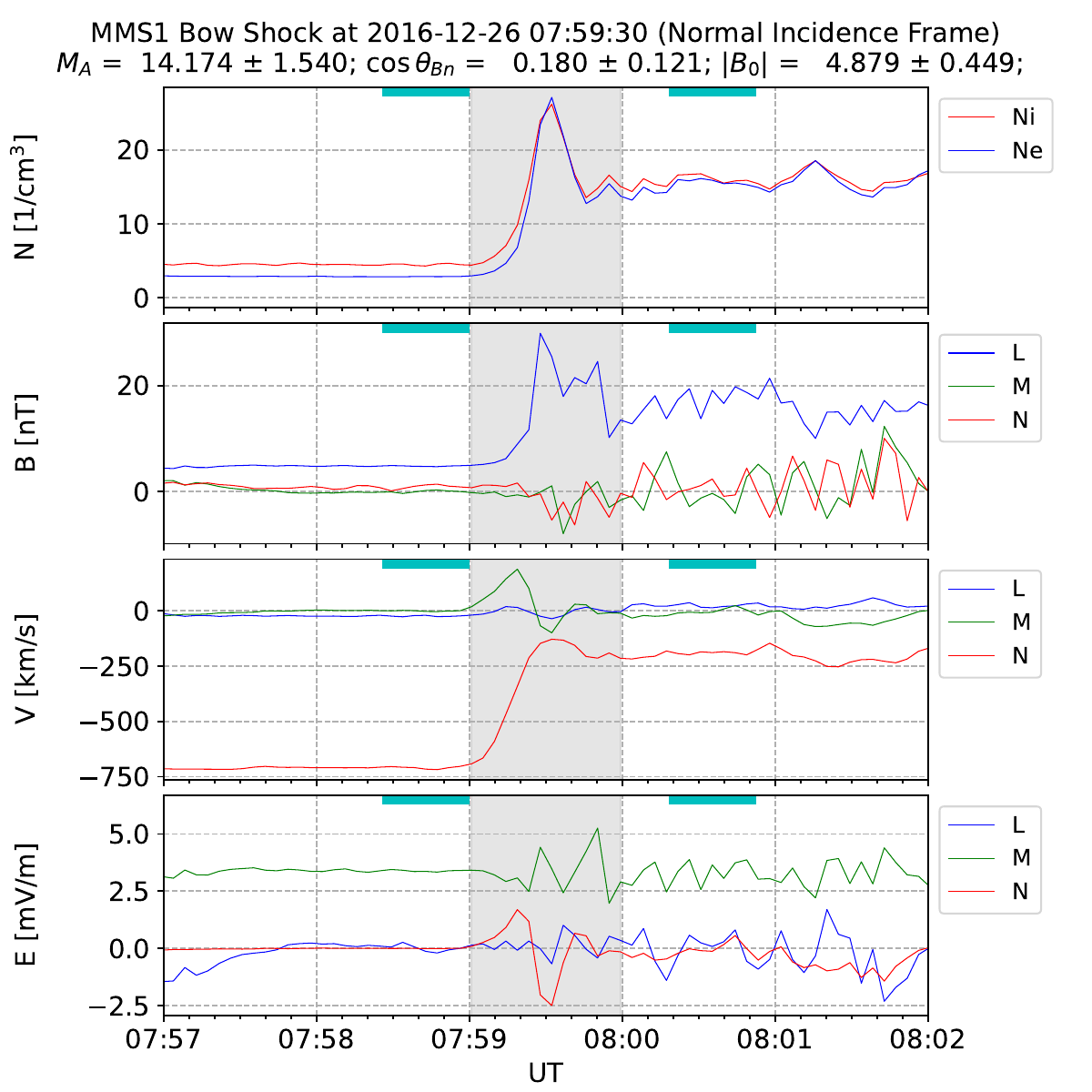}
  \caption{Example of the event observed at around 07:59:30 UT on December 26, 2016 (Event 1). From top to bottom, the ion and electron densities, magnetic field, ion flow velocity, and electric field (defined by $\bm{E} = - \bm{V} \times \bm{B}$) are shown. The vector quantities are represented with the $LMN$ coordinate, and the frame transformation into NIF is performed. The gray area indicates a manually selected shock transition time interval. The cyan bars on top of each panel indicate the upstream (left) and downstream (right) time intervals automatically selected.}
  \label{fig:example1}
\end{figure}

Fig.~\ref{fig:example1} shows an example of a good event (Event 1) observed at around 07:59 UT on December 26, 2016. The expected MHD shock properties are reasonably well satisfied in this event. From top to bottom, the ion and electron densities, magnetic field, ion flow velocity, and electric field (defined by $\bm{E} = - \bm{V} \times \bm{B}$) are shown. The velocity and the electric field are converted into NIF, and all the vector quantities are represented with the $LMN$ coordinate. Note that the averaged magnetic field of $4.5 \, {\rm s}$ time resolution is shown. This event is an inbound shock crossing, which is clear from the density being compressed from the earlier solar wind time interval with a fast plasma flow to the decelerated magnetosheath. The cyan bars on top of each panel indicate the upstream (left) and downstream (right) time intervals automatically selected. We estimated the \Alfven Mach number and the magnetic obliquity as follows: $\MA = 14.2 \pm 1.5$, $\cos \tbn = 0.18 \pm 0.12$. We see that only $B_{l}$ experiences compression across the shock, while $B_{n}$ and $B_{m}$ are nearly constant. Although we theoretically expect $B_{n} = {\rm const.}$ and $B_{m} = 0$, it is difficult to distinguish a small constant and zero for the given large fluctuations, particularly in the downstream, suggesting the importance of checking the electric field. It is clearly seen in the bottom panel that $E_{m}$ is nearly constant, while $E_{l}$ and $E_{n}$ are roughly zero across the shock. The choice of the NIF and the $LMN$ coordinate thus seems reasonable. The velocity normal to the shock $V_{n}$ is the primary component that is decelerated across the shock, which is a natural consequence as both the electric and magnetic fields are consistent with the MHD shock properties.

\begin{figure}[htbp]
  \centering
  \includegraphics[width=0.50\textwidth]{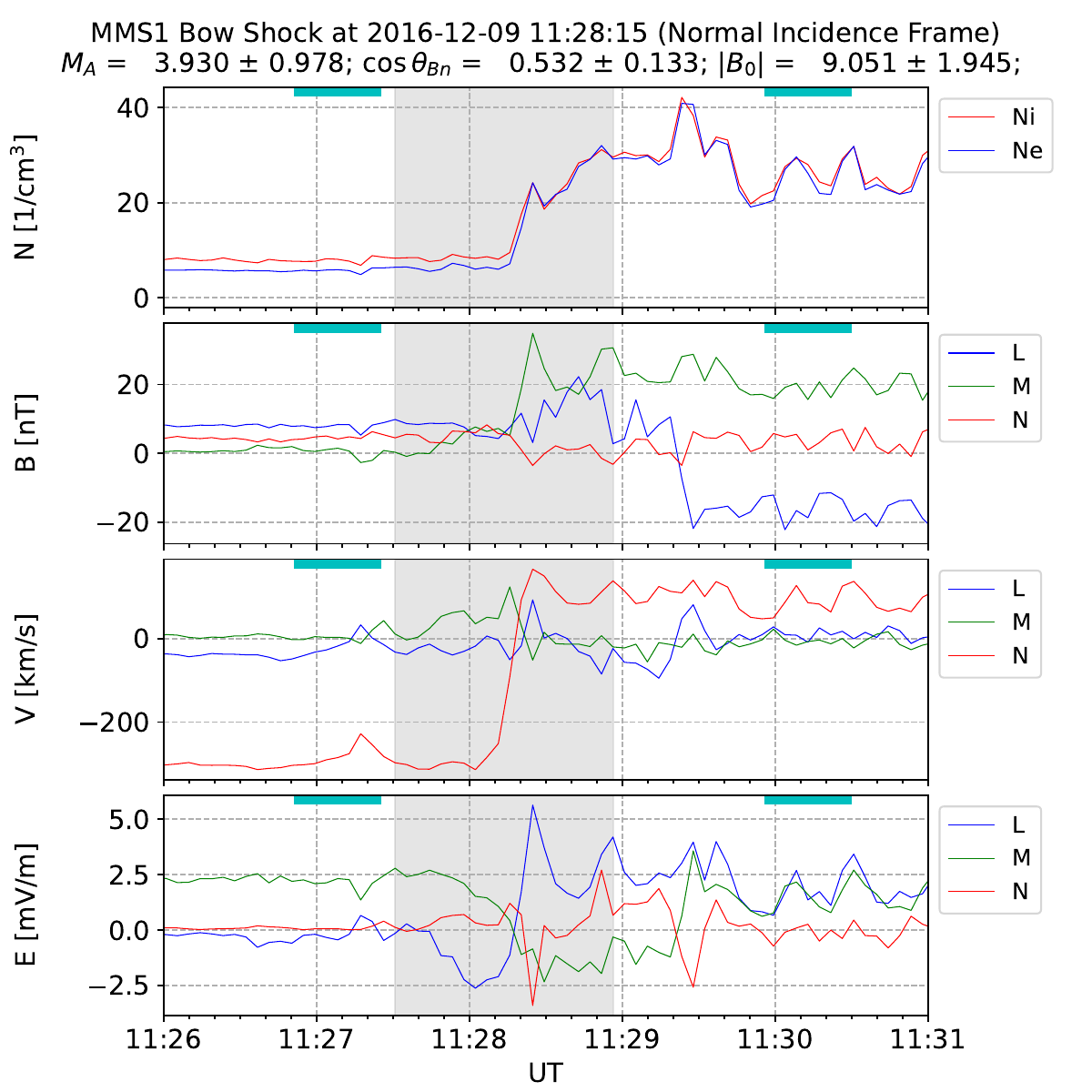}
  \caption{Example of the event observed at around 11:28 UT on December 9, 2016 (Event 2) shown with the same format as Fig.~\ref{fig:example1}.}
  \label{fig:example2}
\end{figure}

Fig.~\ref{fig:example2} presents another example (Event 2) of an inbound shock crossing observed at approximately 11:28 UT on December 9, 2016. The same method, however, gives a result inconsistent with the theoretical expectation. While the plasma flow $V_{n}$ is the primary component of deceleration, both the magnetic and electric fields do not follow the MHD shock properties. This may result from the rotation of the magnetic field at 11:29:20 on the downstream side of the shock. The downstream interval chosen by the algorithm happened to be after the rotation, which is clearly not appropriate. We thus think that the estimated shock parameters are unreliable. Instead of trying to adjust the time interval manually for such cumbersome events, we simply discarded them and collected only cleaner events better suited for the statistical analysis.

\subsection{Whistler Wave Power}
\label{sec:wave}

\subsubsection{Shock Transition Layer}
\label{sec:stl}
It is well known that high-frequency whistler waves (with frequencies higher than $\sim 10$ \% of the local electron cyclotron frequency) are seen not only within STL but also ahead and behind the shock depending on the shock parameters \citep{Zhang1999a,Hull2012a,Wilson2017a,Oka2017a,Shi2020a,Page2021a}. Since the purpose of this paper is to obtain statistical information on the wave power that is relevant for SSDA, we want to minimize the contribution outside STL. To this end, we simply define STL using the magnetic field strength $|\bm{B}|$. As the fast-mode shock is compressional, we may use it as a proxy of the relative position with respect to the shock.

In short, we defined STL as the region satisfying the condition: $B_{\rm min} \leq |B| \leq B_{\rm max}$. The minimum value $B_{\rm min}$ was chosen such that the compression from the unperturbed upstream value $B_0$ exceeds $10 \%$ of the maximum compression. To be more precise, we determined $B_{\rm min}$ by the following equation:
\begin{align}
\frac{B_{\rm min}}{B_0} =
\alpha \left( \frac{B_{\rm max}}{B_0} - 1 \right) + 1,
\end{align}
where $\alpha = 0.1$. This definition depends on $B_{\rm max}$, the maximum magnetic field strength during the shock crossing. In reality, the definition of $B_{\rm max}$ is more complex than one might naively expect due to the presence of multiple peaks and the possibility of contamination by downstream structures. We used the function \texttt{scipy.signal.find\_peaks} in python to find prominent peaks over the entire shock transition. We then chose the one in the most upstream side out of the peaks with values greater than 80\% of the maximum.

With this procedure, we can always systematically determine STL. Notice that this definition does not necessarily give a single time segment of STL. In fact, sometimes, we found multiple non-contiguous segments of STL for a single shock crossing. We think that this is indeed reasonable as the shock structure often exhibits non-stationarity \citep{Leroy1982a,Lembege1992a,Winske1988a,Johlander2016a} or moves in response to the solar wind variations, in which case, the spacecraft may experience multiple excursions of STL. It should also be mentioned that the identification of STL depends on the time resolution of the magnetic field, as the peak location and its height are both dependent on it. We used the averaged magnetic field data with the same time resolution used for the wave power calculation (see below).

\subsubsection{Frequency-Integrated Wave Power}
\label{sec:power}
While we can calculate the power spectral density (PSD) as a function of time during the shock crossing, it is much easier to perform statistical analysis with a scalar quantity that measures the strength of the wave in STL. Therefore, we defined the following frequency-integrated wave power $W$ in units of ${\rm nT}^2$:
\begin{align}
W(f_{\rm min}) \equiv \int_{f_{\rm min}}^{f_{\rm ce}} P(f) df,
\end{align}
where $P(f)$ is the PSD in units of ${\rm nT}^2 / {\rm Hz}$ as a function of frequency $f$ obtained from SCM data. This definition reflects the wave power at the lower bound $f_{\rm min}$ for a PSD rapidly decreasing with frequency, which is usually true, although not always the case. Note that the frequency integration range was defined with respect to the local electron cyclotron frequency $f_{\rm ce}$, which is therefore variable in time.

\begin{figure}[htbp]
  \centering
  \includegraphics[width=0.50\textwidth]{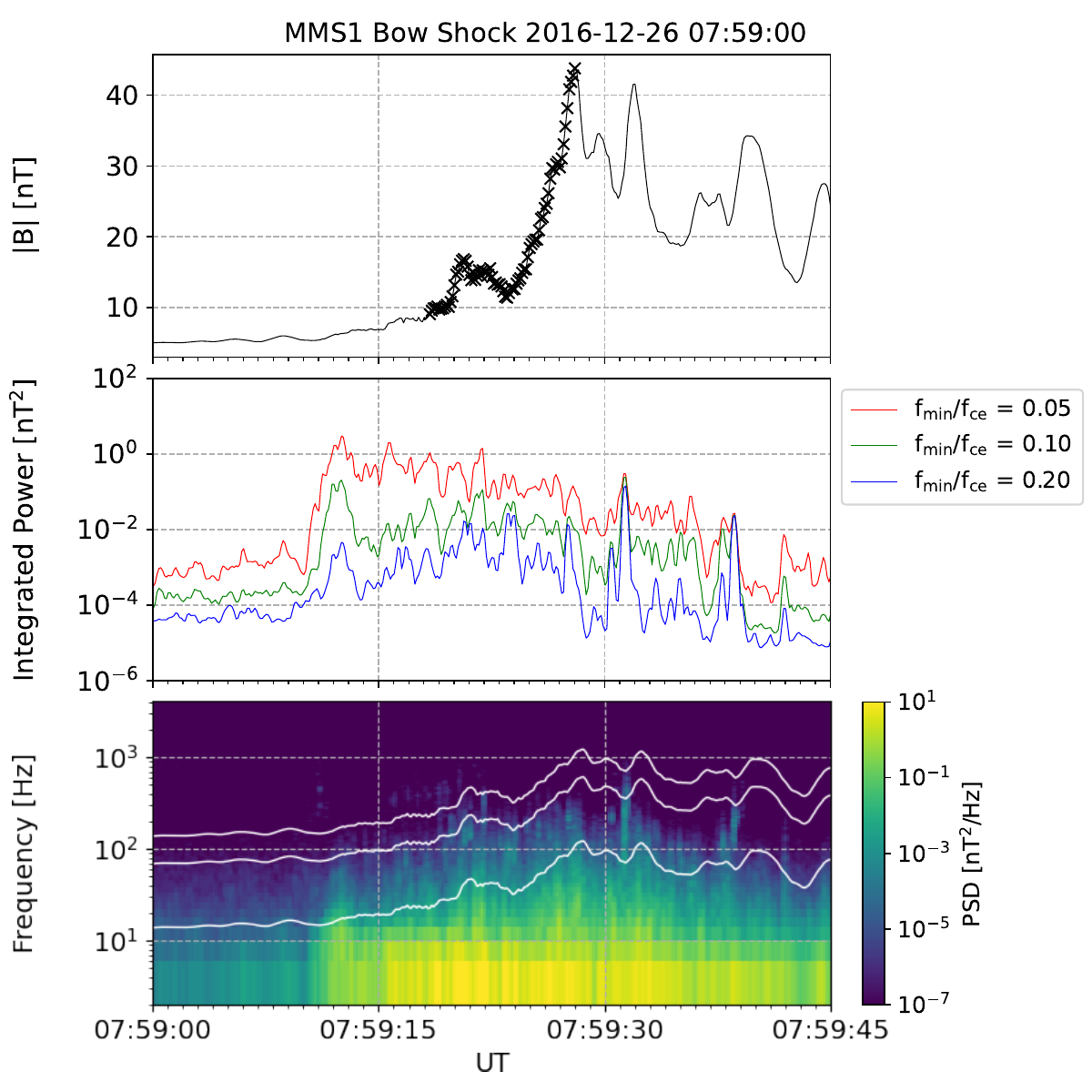}
  \caption{Definition of shock transition layer (STL) and calculation of frequency-integrated wave power. From top to bottom, the magnetic field strength, the frequency-integrated wave power for three lower-bound frequencies $f_{\rm min}/f_{\rm ce} = 0.05, 0.10, 0.20$, and the power spectral density (PSD) are shown. The crosses in the top panel indicate the data points identified as STL. The three white lines in the bottom panel respectively indicate $f/f_{\rm ce} = 0.1, 0.5, 1.0$ for reference.}
  \label{fig:power_timeseries}
\end{figure}

Fig.~\ref{fig:power_timeseries} shows an example of this analysis performed for Event 1. The top panel shows the magnetic field strength. The crosses indicate the data points identified as STL. The middle panel displays the frequency-integrated wave power obtained for three lower-bound frequencies $f_{\rm min}/f_{\rm ce} = 0.05, 0.10, 0.20$, which were obtained by integrating the PSD shown in the bottom panel. We obtained the PSD by performing a short-time discrete Fourier transform on SCM data with the Blackman window of size 2048 data points and an overlap of 1024 data points between the consecutive windows. The resulting time resolution of PSD was $0.25 \, {\rm s}$. We calculated the PSD by taking the sum of the powers of the three magnetic field components. The integration in frequency was performed by using the standard trapezoidal rule. Notice that we did not explicitly try to identify the wave modes, but rather assumed that the PSD is dominated by the whistler waves in the frequency range of interest based on the previous studies \citep{Zhang1999a,Hull2012a,Wilson2017a,Oka2017a,Shi2020a,Page2021a}. Although looking into more detailed wave properties is certainly interesting, we leave it for future work because the wave power is the most important quantity for SSDA.

It is clearly seen that the above procedure identifies STL, where the magnetic field is compressed but weaker than the peak at the overshoot. It has been known that the wave power, particularly at high frequency ($f/f_{\rm ce} \gtrsim 0.1$), is quite variable over very short time scales \citep{Hull2012a,Oka2017a}. While we have not yet understood the origin of such a bursty wave appearance, our primary focus is the average property of the wave power in STL, which we think is the most important quantity to understand the scattering of electrons diffusing around the shock for a sufficiently long time. We thus quantified the wave power in STL by the median value. The uncertainty was evaluated by the first and third quartiles of the distribution and will be shown by error bars in the plots shown in Section \ref{sec:discussion}. We chose the median and quartiles because the wave power is variable by orders of magnitude, and the average and standard deviation do not seem to provide appropriate measures.

We have applied the same procedure automatically for all the events with reasonably well-determined shock parameters and checked the plot with the same format as Fig.~\ref{fig:power_timeseries} and a scatter plot of the wave power as a function of $|B|$ (not shown) for each event. In the following, we mainly discuss various dependence of the frequency-integrated wave power for the three lower-bound frequencies shown here as an example. Note that we only considered $f_{\rm min}/f_{\rm ce} \geq 0.05$ to minimize the impact of the Doppler correction ignored in this study. This limitation, however, does not hinder our objective of quantifying the high-frequency whistler wave power.

\section{Result and Discussion}
\label{sec:discussion}

\subsection{Summary of Event Distribution}
\label{sec:event_distribution}
As a result of the event selection, the number of events suitable for the statistical analysis was reduced to 91 events from 121 candidates. Although we applied the same procedure to all four MMS spacecraft, the results were often nearly identical because of the small spacecraft separation. We checked the consistency of the estimated shock parameters independently determined with the four spacecraft and discarded inconsistent events. We will then discuss the results obtained by MMS1 in the following.

\begin{figure}[htbp]
  \centering
  \includegraphics[width=0.50\textwidth]{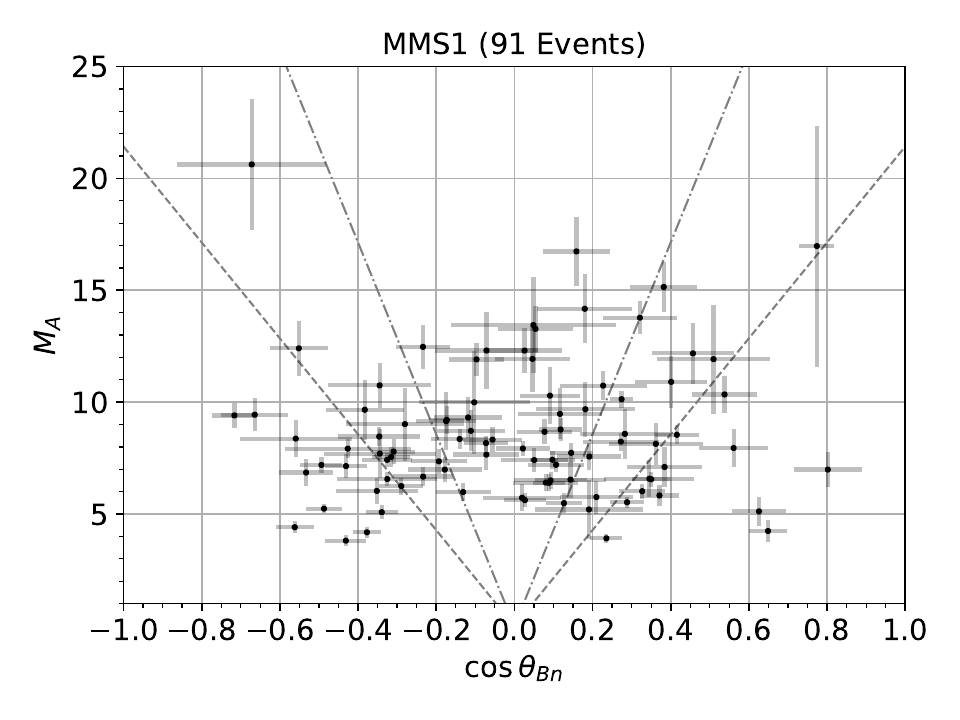}
  \caption{Scatter plot of the estimated parameters $\MA$ and $\cos \tbn$. The dashed and dash-dotted lines respectively indicate $\MA = 1/2 \sqrt{m_i/m_e} \cos \tbn$ and $\MA = \sqrt{m_i/m_e} \cos \tbn$ for reference.}
  \label{fig:event_summary}
\end{figure}

Figure~\ref{fig:event_summary} presents a scatter plot of the estimated parameters $\MA$ and $\cos \tbn$. The dashed line represents the whistler critical Mach number $\MA = (\sqrt{m_i/m_e} \cos \tbn) / 2$ calculated based on the whistler wave phase velocity. Similarly, the dash-dotted line represents twice the whistler critical Mach number $\MA = \sqrt{m_i/m_e} \cos \tbn$. These are shown for reference for reasons that will become clear later in Section \ref{sec:theoretical_threshold}.

We see that the majority of events had Alfven Mach numbers in the range $5 \lesssim \MA \lesssim 15$, which is consistent with the average property of the solar wind. We can also see that the distribution of $\cos \tbn$ is sparse at the quasi-parallel region $|\cos \tbn| \gtrsim 0.7$ because of the event selection bias. We think that the bias does not significantly affect our discussion as long as we focus on the wave power dependence in relation to electron acceleration because quasi-perpendicular shocks are known to be more efficient in producing energetic electrons, at least at Earth's bow shock\citep{Oka2006a}.

\begin{figure}[htbp]
  \centering
  \includegraphics[width=0.50\textwidth]{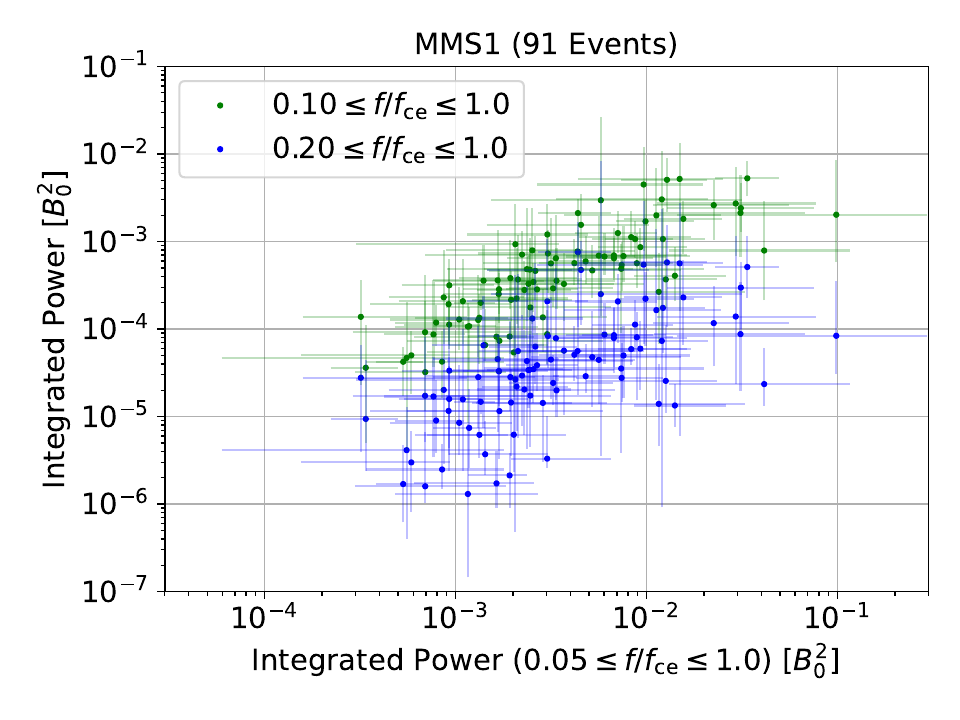}
  \caption{Relation between integrated wave powers of different frequency bands. The green and blue symbols respectively indicate the integrated wave power in the middle ($0.10 \leq f/f_{\rm ce} \leq 1.0$) and high ($0.20 \leq f/f_{\rm ce} \leq 1.0$) frequency band as a function of the lowest frequency band ($0.05 \leq f/f_{\rm ce} \leq 1.0$).}
  \label{fig:power_summary}
\end{figure}

Fig.~\ref{fig:power_summary} shows the relation between the obtained frequency-integrated wave power normalized to $B_0^2$ for the three frequency bands. It is clear that the high-frequency powers positively correlate with the power in the lowest band. The result indicates that the ratios between different frequencies are roughly constant in a statistical sense, meaning that the PSD is approximately represented by a power law in frequency. We should note that, however, high-frequency whistler waves look coherent and appear sporadically in time, and the instantaneous PSD does not necessarily follow the simple power law (as can also be seen in Fig.~\ref{fig:power_timeseries}). We thus believe that the generation of high-frequency whistler waves is not likely a result of the classical forward turbulent cascade. Nonetheless, the approximate power-law form of PSD, on average, may be useful in the theoretical modeling of particle acceleration.

In the following, we will further restrict the dataset to minimize the impact of errors in the estimated parameters without losing statistics. It is easy to anticipate that the parameter determination for quasi-parallel shocks is less accurate because of the more turbulent shock transition. In addition, finding $\tbn$ dependence appears to be difficult at nearly perpendicular shocks where the value of $\cos \tbn$ becomes comparable to or even smaller than the typical errors $\sigma \left( \cos \tbn \right) \sim 0.1$ (see, Fig.~{\ref{fig:event_summary}}). This error becomes particularly significant when we consider the dependence on $\MA/\cos \tbn$ (for which the error is significantly amplified). We will thus discuss the dataset in the range $|\cos 80\degree| < |\cos \tbn| < |\cos 60\degree|$. The number of events in this range is 45, which we think is still sufficient for the statistical analysis. In the rest of this section, we will show plots for these restricted events. For completeness, corresponding plots for all the events will be shown in Appendix \ref{sec:appendix_all}.

\subsection{\Alfven Mach Number Dependence}
\label{sec:Ma_dependence}
We have searched for possible parameter dependence of the frequency-integrated wave power on various parameters, including $\MA$, $\tbn$, $\beta_{\rm omni}$. No significant correlation with $\tbn$ has been found. There might be some $\beta_{\rm omni}$ dependence. However, the most significant correlation was found with $\MA$ as we see below.

Fig.~\ref{fig:mach_nif_selected} shows the dependence of the frequency-integrated power on the \Alfven Mach number $\MA$. From left to right, scatter plots for the three different lower-bound frequencies are shown: $f_{\rm min}/f_{\rm ce} = 0.05$ (left), $f_{\rm min}/f_{\rm ce} = 0.10$ (middle), $f_{\rm min}/f_{\rm ce} = 0.20$ (right), respectively. The red and blue symbols indicate events with $\beta_{\rm omni}$ higher and lower than the median value, respectively. We have made this distinction of high and low $\beta$ events because the solar wind \Alfven Mach number tends to be higher in high $\beta$ periods (i.e., \Alfven Mach number is primarily controlled by the magnetic field strength, and the temperature is nearly constant). This indicates that the dependence on $\MA$ and $\beta$ may potentially be degenerated. However, by dividing the events into two groups, we can see that both groups have similar dependence on $\MA$. The power increases with increasing $\MA$ at lower Mach numbers $\MA \lesssim 10$ and then appears to saturate at higher Mach numbers $\MA \gtrsim 10$. Therefore, we think that the trend is indeed $\MA$ dependence rather than $\beta$ dependence.

It seems intuitive that higher \Alfven Mach numbers would result in larger normalized wave amplitudes due to the increased free energy available for wave generation. However, it is worth noting that the increase in wave power appears to be more rapid than the $\propto \MA^2$ dependence expected based on the simple energetics argument, indicating that the wave generation efficiency itself increases. Assuming that the wave generation in STL is a result of linear plasma instabilities, the apparent saturation observed at higher Mach numbers could be attributed to nonlinear effects. The saturated wave power at the lowest frequency reaches up to $\sim 10^{-2} B_0^2$. Considering the magnetic field compression in STL, this corresponds to a wave amplitude that is a few percent of the local magnetic field strength. Nonlinear effects might start to play a role to cause the saturation at this amplitude level. Nevertheless, the sparsity of the data points at higher Mach numbers does not allow us to draw a firm conclusion.

\begin{figure*}[htbp]
  \centering
  \includegraphics[width=1.0\textwidth]{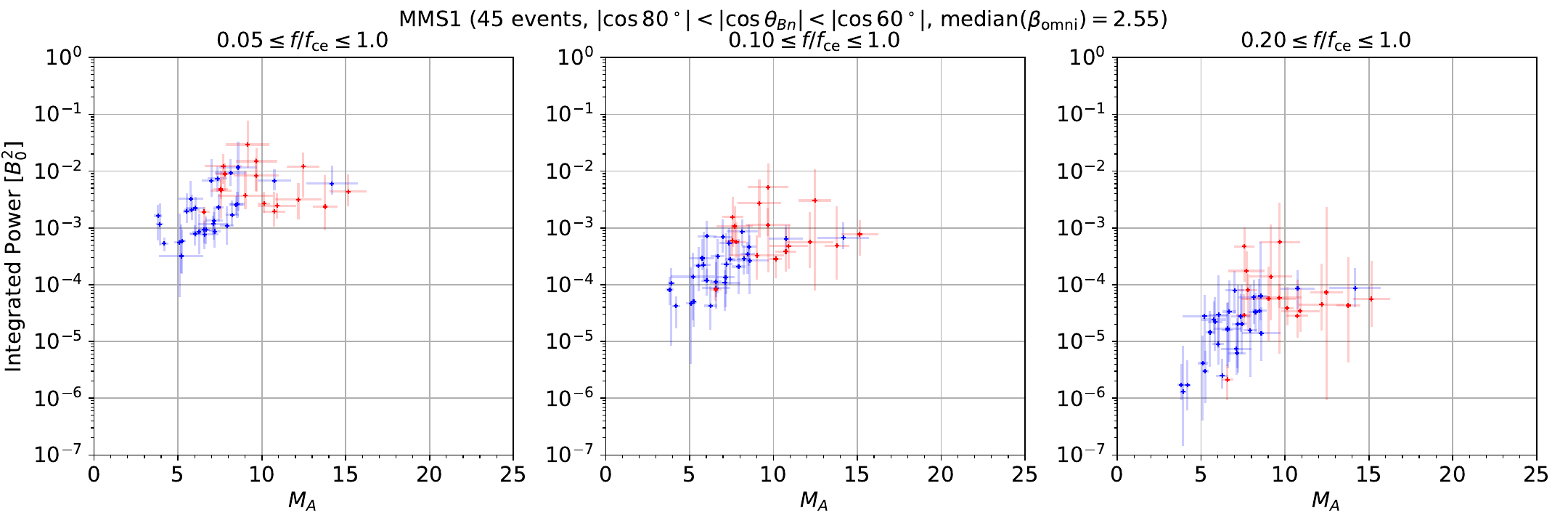}
  \caption{Relation between integrated wave power and \Alfven Mach number $\MA$. From left to right, scatter plots for the three different lower-bound frequencies are shown: $f_{\rm min}/f_{\rm ce} = 0.05$ (left), $f_{\rm min}/f_{\rm ce} = 0.10$ (middle), $f_{\rm min}/f_{\rm ce} = 0.20$ (right), respectively. The red and blue symbols indicate events with $\beta_{\rm omni}$ higher and lower than the median value.}
  \label{fig:mach_nif_selected}
\end{figure*}

\subsection{Relation to Theoretical Threshold}
\label{sec:theoretical_threshold}
\begin{figure*}[htbp]
  \centering
  \includegraphics[width=1.0\textwidth]{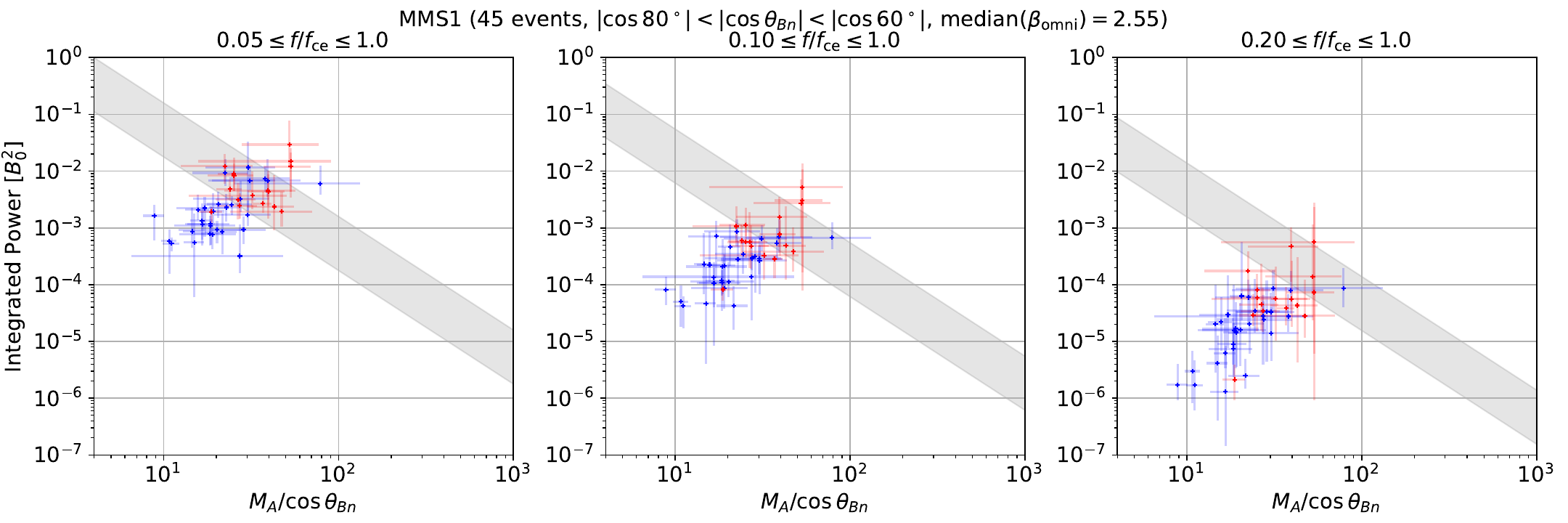}
  \caption{Relation between integrated wave power and \Alfven Mach number in the de Hoffmann-Teller frame $\MAHTF = \MA / \cos \tbn$. The format is similar to Fig.~\ref{fig:mach_nif_selected}. The gray area indicates the theoretical threshold (with an order of magnitude uncertainty taken into account) for efficient electron acceleration by SSDA.}
  \label{fig:mach_htf_selected}
\end{figure*}

As we mentioned already, we have not been able to find a significant correlation between the wave power and the shock obliquity $\tbn$. On the other hand, a rather clear correlation was found with $\MA/\cos\tbn$ as shown in Fig.~\ref{fig:mach_htf_selected}. Furthermore, this dependence is useful for direct comparison with the theory of SSDA.

The theory indicates that the efficiency of electron acceleration for a given pitch-angle scattering rate is a strong function of $\MAHTF \equiv \MA/\cos\tbn$, which is physically understood as the \Alfven Mach number in the de Hoffmann-Teller frame (HTF). Specifically, for an electron to be accelerated by SSDA, the scattering rate must exceed a theoretical threshold value that is proportional to the electron's energy and $(\MAHTF)^{-2}$. The scattering rate should be related to the wave power in STL using the quasi-linear theory, whereas the electron's energy should be related to the wave frequency through the cyclotron resonance condition. Based on this, the threshold suitable for comparison with the frequency-integrated wave power can be written as (see, Appendix \ref{sec:appdendix_threshold}):
\begin{align}
  \frac{W^{*} (f_{\rm min})}{B_0^2}
  & = \frac{2}{3 \pi \eta}
  \left( \frac{B_{\rm STL}}{B_0} \right)
  \left( \MAHTF \right)^{-2}
  \int_{k_{\rm min}}^{\infty}
  \left( \frac{v}{\VAe} \right)^3
  \frac{dk_{\rm res}}{\Wce/\VAe},
  \label{eq:threshold}
\end{align}
where $B_{\rm STL}$ is the magnetic field strength in STL and $\eta$ is the thickness of STL in units of the ion gyroradius defined with the upstream flow velocity and the magnetic field strength. The particle velocity $v$ is normalized to the electron \Alfven speed $\VAe$, and the resonant wavenumber $k_{\rm res}$ is normalized to the reciprocal of the electron skin depth $\VAe/\Wce$ with $\Wce$ being the electron cyclotron frequency. The relation between the velocity and the resonant wavenumber is determined by the cyclotron resonance condition and the cold plasma dispersion relation for parallel propagation. The lower bound for the integration $k_{\rm min}$ is related to $f_{\rm min}$, whereas the upper bound becomes infinity because $k \rightarrow \infty$ for $f \rightarrow f_{\rm ce}$ with the cold plasma dispersion relation.

We should mention that the theoretical threshold is an order of magnitude estimate as, for instance, the quasi-linear theory may not always be applicable, and the resonance condition considers only parallel propagation and ignores the pitch-angle dependence. Therefore, the threshold is shown with the gray area in Fig.~\ref{fig:mach_htf_selected} representing uncertainty by a factor of three larger than and smaller than the fiducial value calculated by using typical values for STL \citep{Amano2020a}: $B_{\rm STL}/B_0 = 2$ and $\eta = 0.5$. We note that the integration in wavenumber is of order unity and roughly determined by the resonant particle velocity $v$ corresponding to the lower-bound frequency $f_{\rm min}$, which explains the decreasing trend as frequency increases.

The large scatter of the data points and the uncertainty in the theoretical threshold do not allow us to precisely determine the critical point beyond which the wave power exceeds the threshold. Nevertheless, with the increasing trend of wave power as a function of $\MAHTF$ and the theoretical dependence $(\MAHTF)^{-2}$ of the threshold given by Eq.~\ref{eq:threshold}, it is clear that, at some point, the waves become strong enough to sustain efficient particle acceleration by SSDA.

The resonant particle energies can be roughly estimated using typical parameters in STL ($B = 10 \, {\rm nT}$, $n = 10 \, {\rm cm^{-3}}$) as $\sim 60 \, {\rm eV} \, (f_{\rm min}/f_{\rm ce} = 0.20)$, $\sim 200 \, {\rm eV} \, (f_{\rm min}/f_{\rm ce} = 0.10)$, and $\sim 400 \, {\rm eV} \, (f_{\rm min}/f_{\rm ce} = 0.05)$, respectively. Observations of the electron energy spectrum have shown that the power-law tail in the spectrum typically forms beyond $\sim 100 \, {\rm eV}$ \citep{Feldman1982a,Feldman1983b,Schwartz1988a}. The resonant energy for the highest frequency whistlers thus appears to be smaller than the lowest energy in the power-law tail. Considering that the resonant energy is the minimum energy required for cyclotron resonance, we think that the highest frequency whistlers interact with electrons at the transition energy between the thermal and non-thermal populations. Although the SSDA theory ignores the effect of electrostatic waves, it is possible that they play a role in the electron energization at the transition energy and below. Indeed, intense electrostatic waves observed in STL can have potentials as high as a few tens of ${\rm eV}$ \citep{Vasko2022a,Sun2022a,Wang2023a}, indicating that they may contribute to the heating of low-energy electrons. On the other hand, since the electrons in the non-thermal tail have much larger energies than the wave electrostatic potential, we think that they are primarily scattered by the whistlers with $f/f_{\rm ce} \lesssim 0.1$. Looking at the two panels for lower-frequency bands in Fig.~\ref{fig:mach_htf_selected}, we estimate that a shock of $\MAHTF \gtrsim 30-60$ will be an efficient accelerator of electrons in the non-thermal tail with energies $\gtrsim 100 \, {\rm eV}$.

This result is in remarkable agreement with the one reported by \citet{Oka2006a}, who found that the spectral index of energetic electrons in Earth's bow shock is regulated by $\MAHTF$. Figure 4 of \citet{Oka2006a} indicates that the spectral index becomes harder when $\MAHTF$ exceeds a certain critical value, and otherwise, there is no systematic trend. We read this critical value roughly as $M_{\rm A}^{{\rm HTF}*} \sim 2 \sqrt{(27/64) (m_i/m_e)} \approx 56$, where the factor inside the square root comes from the definition of the whistler critical Mach number based on the whistler wave group velocity. We can therefore interpret the result as follows: electron acceleration by SSDA is triggered to produce the power-law tail at energies $\gtrsim 100 \, {\rm eV}$ when $\MAHTF$ exceeds the critical value $\sim 30-60$ because the wave power for $f/f_{\rm ce} \lesssim 0.1$ becomes large enough to efficiently confine electrons within STL. The hard spectral index in the super-critical regime is consistent with the theoretical prediction \citep{Amano2022b}, whereas the scatter observed in the sub-critical regime may reflect an intrinsic scatter of the suprathermal electron power-law tail in the solar wind.

\subsection{Wave Generation Mechanism}
\label{sec:wave_generation}
Our primary objective in this paper is to establish an empirical relationship between the wave power and various shock parameters. Given the success, particularly in relation to SSDA, it is interesting to extend the discussion to potential wave generation mechanisms.

\begin{figure*}[htbp]
  \centering
  \includegraphics[width=1.0\textwidth]{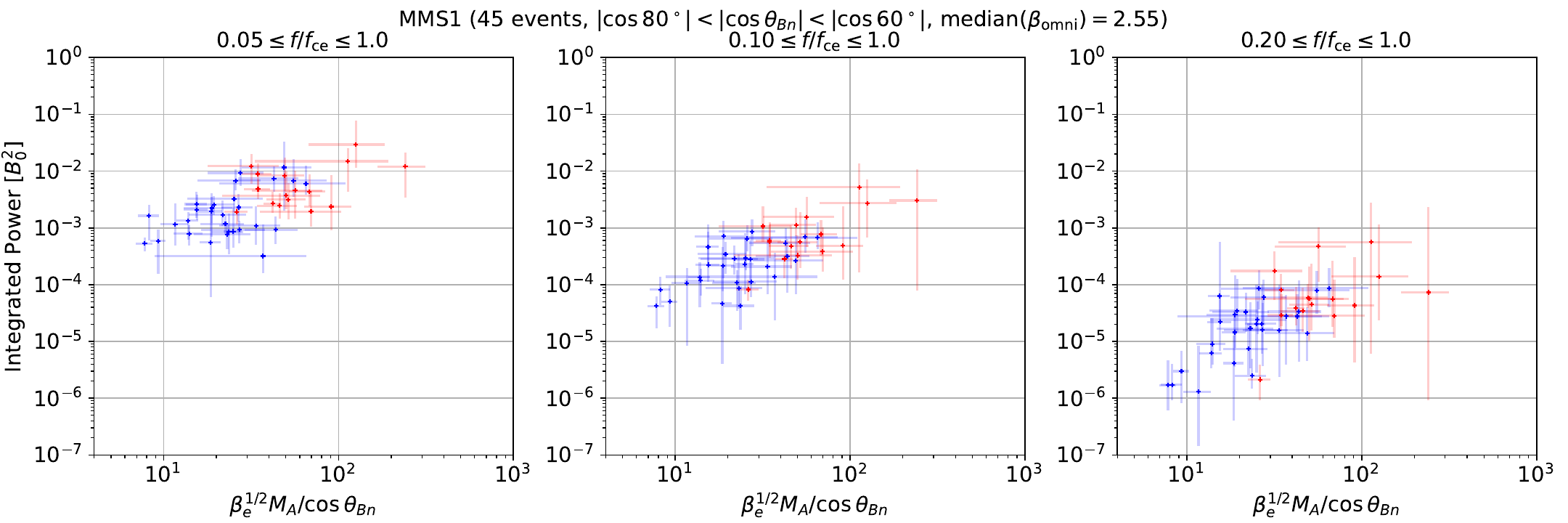}
  \caption{Relation between integrated wave power and $\beta_e^{1/2} \MAHTF = \beta_e^{1/2} \MA / \cos \tbn$. The format is similar to Fig.~\ref{fig:mach_nif_selected}.}
  \label{fig:mach_beta_selected}
\end{figure*}

Although the critical value of $\MAHTF$ discussed in the previous section nearly coincides with the whistler critical Mach number, there does not seem to be a causal relationship between the two. Recall that the whistler critical Mach number is defined as the critical point beyond which whistler waves, even at the maximum phase or group velocity, cannot propagate upstream. Therefore, it does not predict the wave generation itself. It is natural to expect that dispersive whistler waves will be emitted associated with the steepening of the shock front. However, it would not be easy to imagine this mechanism produces coherent high-frequency ($f/f_{\rm ce} \sim 0.1$) whistler waves with a bursty appearance, as reported previously. We thus think that the whistler waves of our interest are most likely generated by some sort of plasma instabilities. Nevertheless, it is important to mention that once the waves are generated in STL, they cannot escape upstream regardless of the propagation direction at a whistler-super-critical shock. Even in the sub-critical regime, higher $\MAHTF$ shocks tend to accumulate the whistler wave energy in STL. This is one of the possible reasons to explain the observed $\MAHTF$ dependence of the wave power (Fig.~\ref{fig:mach_htf_selected}).

One of the most well-known mechanisms of generating quasi-parallel whistler waves is the so-called whistler instability \cite{Kennel1966a} arising from the perpendicular electron temperature anisotropy $T_{\perp} / T_{\parallel} > 1$. Since the thermal electrons convected into the shock will be adiabatically heated in the perpendicular direction by the shock-compressed magnetic field, it is natural to expect that the favorable temperature anisotropy will develop. However, since this mechanism in its simplest form does not explain a dependence on $\MAHTF$, some modifications, such as the presence of a beam \cite{Tokar1984a}, must be necessary to explain the observed dependence.

\citet{Amano2022b} discussed possible instabilities for high-frequency whistler wave generation that have been considered in the past. Interestingly, a lot of instabilities have a dependence on $\MAHTF$. Among these, the modified-two-stream (MTS) instability \citep{Matsukiyo2003a,Matsukiyo2006a} driven either by the incoming or reflected ion beams requires lower values $\MAHTF \lesssim \left( m_i/m_e \right)^{1/2}$. Although we do not completely rule out the possibility of MTS, the observed trend appears to be inconsistent. This may not be surprising in that MTS predicts wave generation at much lower frequencies. On the other hand, a quasi-parallel whistler wave generation mechanism of \citet{Amano2010a} driven by an electron beam with loss-cone produced by the adiabatic SDA favors $\MAHTF \gtrsim \left( m_i/m_e \right)^{1/2} \beta_e^{1/2}$. This mechanism is consistent with the observations in that the high-frequency whistlers is expected and the propagation direction is nearly parallel (or anti-parallel) to the ambient magnetic field \citep{Hull2012a,Oka2017a}. A similar dependence on $\MAHTF$ has been suggested by \citet{Levinson1992a,Levinson1996a} for oblique whistler wave generation $\MAHTF \gtrsim \left( m_i/m_e \right)^{1/2} \beta_e^{-1/2}$. While this mechanism requires a finite wave normal angle, it may better fit into the context because it considers waves driven by an upstream directed flux of electrons diffusively confined in STL, which is exactly the situation considered in SSDA. Although no explicit analysis has been given so far, oblique whistlers reported previously \citep{Oka2019a,Hull2020a} may potentially be understood with this wave excitation mechanism.

With this theoretical background in mind, we have searched for the possible dependence on $\beta_e^{1/2} \MAHTF$ and $\beta_e^{-1/2} \MAHTF$ and found that the former appears to have a better correlation with the wave power, which is shown in Fig.~\ref{fig:mach_beta_selected}. We may consider that the apparent correlation favors the oblique whistler wave generation mechanism proposed by \citet{Levinson1992a,Levinson1996a}. We note that, however, the theory predicts a threshold value for the wave generation rather than the continuously increasing wave power as observed. In addition, although we have not distinguished the wave propagation direction, both parallel and anti-parallel propagating waves (or propagation toward the upstream and the downstream) have been reported in the literature \citep{Oka2017a,Amano2020a}, which is not consistent with naive expectation of the theory. Therefore, we think that the correlation found here alone is not necessarily conclusive evidence for the oblique whistler wave generation mechanism.

It is, nevertheless, worth noting that if the $\beta_e$ dependence is confirmed, it has significant implications for astrophysical applications. Shock acceleration of electrons in high-$\beta$ medium has been a topic of substantial interest in the context of galaxy clusters \citep{Matsukiyo2011a,Guo2014a,Guo2014b}. Although, in general, shock waves driven by mergers of galaxy clusters are weak with typical sonic Mach numbers of a few, the high plasma beta $\beta \gg 1$ may potentially make the shock an efficient electron accelerator if the shock is sufficiently oblique. Another example is the so-called termination shock of the solar flare \citep{Mann2009a,Guo2012a,Chen2015a}. In the standard magnetic reconnection scenario of the solar flare, the reconnection outflow is decelerated by a standing termination shock, which is likely to be quasi-perpendicular. Since the upstream region corresponds to the reconnection outflow, the pre-shock medium will already be in a high-$\beta$ state. Understanding the $\beta_e$ dependence will shed light on the electron acceleration efficiency in these astrophysical environments where in-situ observations are not available.

It is fair to mention that the possible wave generation mechanisms favoring high $\MAHTF$ discussed above have theoretical imperfection. The quasi-parallel whistler wave generation of \citet{Amano2010a} assumes a constant beam density, which is, however, likely to be a function of $\beta_{e}$. If the beam is generated by SDA, a higher $\beta_{e}$ shock will produce a higher beam density that may compensate for the predicted $\beta_{e}$ dependence. On the other hand, the oblique whistler wave generation by \citet{Levinson1992a,Levinson1996a} ignores the effect of Landau damping by the background electrons. Essentially the same oblique whistler generation mechanism has been discussed recently in other contexts \citep{Krafft2010a,Roberg-Clark2018a,Verscharen2019a,Vasko2019a}, some of which pointed out the importance of the electron Landau damping. More detailed investigation should be carried out to confirm if the instability persists in high-$\beta_{e}$ conditions where the Landau damping is significant. Similarly, the data analysis presented in this paper does not distinguish the wave propagation direction. In principle, a more detailed analysis of the wave property may provide information to identify the wave generation mechanism. Further theoretical and observational studies are both needed for a more comprehensive understanding of electron acceleration at a shock in a wider range of parameter space.

\section{Summary and Conclusion}
\label{sec:conclusion}
This study has presented the statistical analysis of high-frequency whistler waves observed in the shock transition layer (STL) of Earth's bow shock using the data obtained by MMS spacecraft. We have found correlations between the frequency-integrated whistler wave power in STL and shock parameters. In general, high \Alfven Mach number shocks tend to have larger wave power. While the shock obliquity $\tbn$ alone does not seem to significantly control the wave power, a significant correlation with the de Hoffmann-Teller frame \Alfven Mach number $\MAHTF = \MA/\cos\tbn$ has been found. This allows us to compare the observation with the theory of stochastic shock drift acceleration (SSDA). We have found that the wave power exceeds the threshold predicted by the SSDA theory when $\MAHTF \gtrsim 30-60$, indicating that efficient electron acceleration is realized in this regime. This is very well consistent with the previous statistical study by \citet{Oka2006a} on the spectral index of energetic electrons at Earth's bow shock. We have also discussed the possible wave generation mechanism based on the statistical analysis. Although we have not been able to identify the mechanism, the statistical analysis suggests that high-$\beta_e$ shocks may be favorable for the wave generation and, thus, electron acceleration.

In conclusion, we consider that this paper provides further support for the validity of SSDA as the mechanism for electron acceleration at the quasi-perpendicular Earth's bow shock. Considering the theoretical scaling of the maximum achievable energy $\propto (\MAHTF)^2$, we think that the electron injection scenario by SSDA at higher Mach number astrophysical shocks has become even more promising. Nevertheless, we should keep in mind that the high-frequency whistler wave is the crucial element for the \textit{initiation} of SSDA at slightly above the thermal energy. Scattering and acceleration of higher energy electrons require lower frequency fluctuations, although this is expected to be less problematic as the wave power at lower frequencies is much higher in general. On the other hand, we think that the heating of electrons at and below the thermal energy is likely to be dominated by electrostatic waves. The efficiency of heating probably determines the thermal and non-thermal transition energy and the flux of electrons injected into SSDA. A more comprehensive understanding of electron acceleration at shocks thus requires further investigation of the activity of both electrostatic and electromagnetic waves for a wide range of frequencies.

\begin{acknowledgments}
We thank the MMS team for providing the data used in this study. This work was supported by JSPS KAKENHI grant No.~22K03697 and the bilateral joint research project JPJSBP120233504. T.~A. acknowledges support by the International Space Science Institute (ISSI) in Bern through ISSI International Team project \#520 “Energy Partition across Collisionless Shocks”.
\end{acknowledgments}

\section*{data availability}
All the MMS data used in this study are publicly available at the MMS Science Data Center (https://lasp.colorado.edu/mms/sdc/public/). Analysis codes used in this study and the list of bow shock crossings including the determined parameters are available at \url{https://doi.org/10.5281/zenodo.10457321}.

\appendix
\section{Scatter Plot for All Events}
\label{sec:appendix_all}
\begin{figure*}[htbp]
  \centering
  \includegraphics[width=1.0\textwidth]{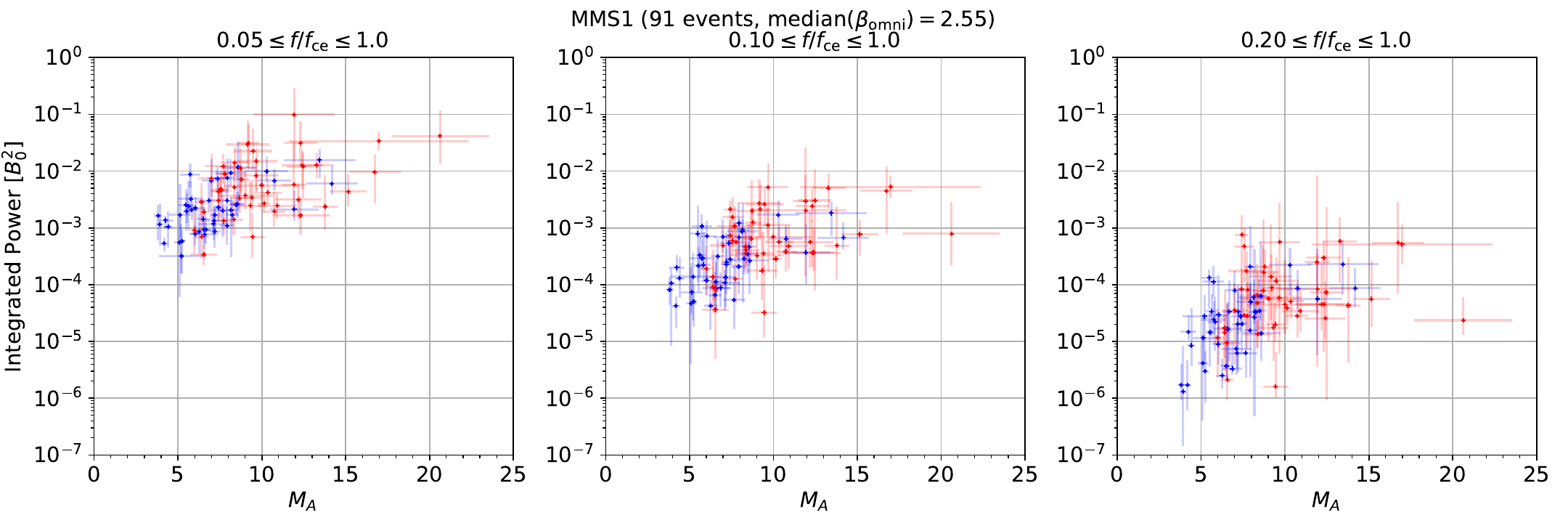}
  \caption{Relation between integrated wave power and \Alfven Mach number $\MA$ for all the event without restriction on $\tbn$. The format is the same as Fig.~\ref{fig:mach_nif_selected}.}
  \label{fig:mach_nif_all}
\end{figure*}

\begin{figure*}[htbp]
  \centering
  \includegraphics[width=1.0\textwidth]{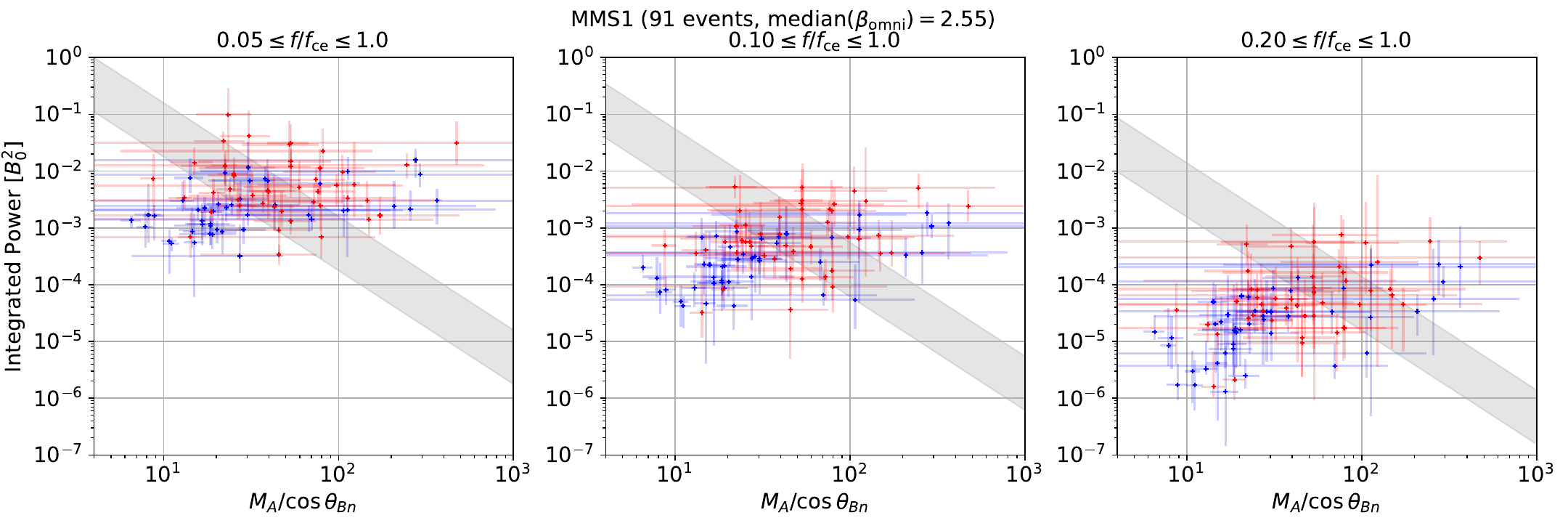}
  \caption{Relation between integrated wave power and \Alfven Mach number in the de Hoffmann-Teller frame $\MAHTF = \MA / \cos \tbn$ for all the event without restriction on $\tbn$. The format is the same as Fig.~\ref{fig:mach_htf_selected}.}
  \label{fig:mach_htf_all}
\end{figure*}

\begin{figure*}[htbp]
  \centering
  \includegraphics[width=1.0\textwidth]{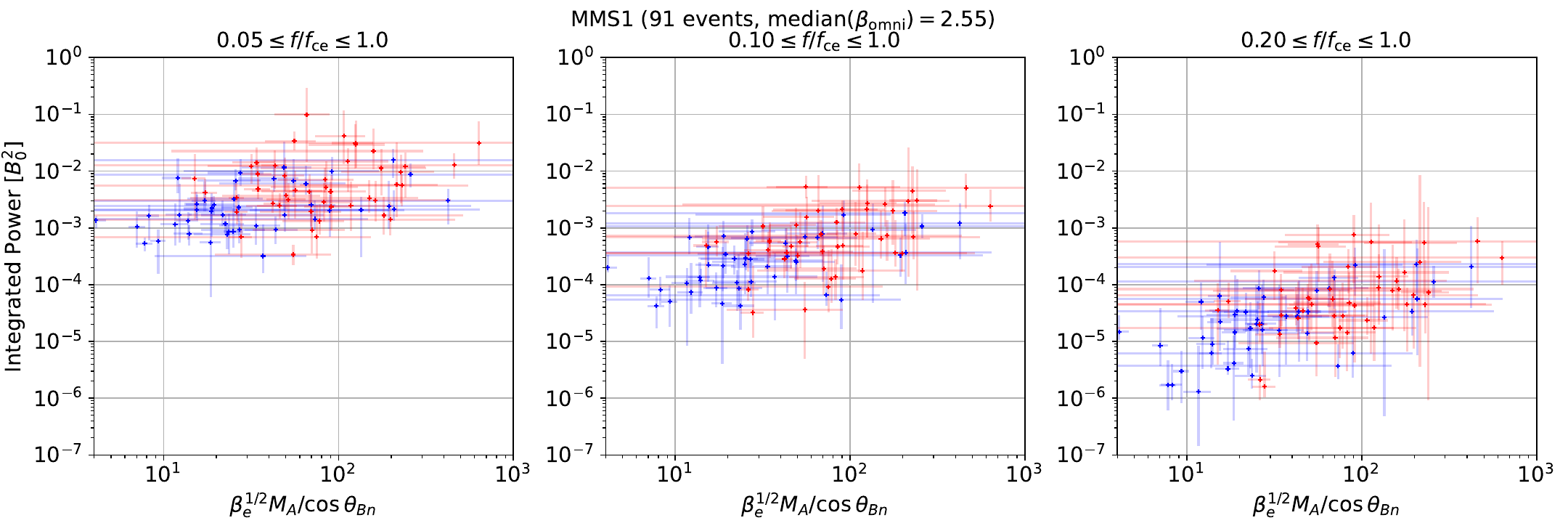}
  \caption{Relation between integrated wave power and $\beta_e^{1/2} \MAHTF = \beta_e^{1/2} \MA / \cos \tbn$ for all the event without restriction on $\tbn$. The format is the same as Fig.~\ref{fig:mach_beta_selected}.}
  \label{fig:mach_beta_all}
\end{figure*}

Plots for all the events without restriction on the shock obliquity $\tbn$ are shown in this appendix. The dependence on $\MA$, $\MA / \cos \tbn$, $\beta_e^{1/2} \MA / \cos \tbn$ are shown in Figs.\ref{fig:mach_nif_all}, \ref{fig:mach_htf_all}, and \ref{fig:mach_beta_all} with the same format as Figs.~\ref{fig:mach_nif_selected}, \ref{fig:mach_htf_selected}, and \ref{fig:mach_beta_selected}, respectively.

The error bars in Fig.\ref{fig:mach_nif_all} are about the same level as those in Fig.\ref{fig:mach_nif_selected}, and the tendency is similar. We see indeed that the wave power tends to saturate at $\MA \gtrsim 10$, consistent with what is seen in Fig.\ref{fig:mach_nif_selected}.

On the other hand, we see significantly larger errors in Figs.\ref{fig:mach_htf_all} and \ref{fig:mach_beta_all} because the errors in the shock obliquity are amplified through the factor $1/\cos \tbn$. Nevertheless, comparison with the restricted dataset suggests similar increasing trends of the wave power as functions of $\MAHTF$ and $\beta_e^{1/2} \MAHTF$. The result might also suggest that there is saturation at high $\MAHTF$ and $\beta_e^{1/2} \MAHTF$. However, we think that it is not possible to draw any firm conclusions due to the large scatter and the errors in the dataset.

\section{Threshold Wave Power}
\label{sec:appdendix_threshold}
For efficient particle acceleration by SSDA, accelerated electrons must be scattered in pitch angle frequently enough. This isotropizes the pitch-angle distribution so that the spatial transport along the magnetic field line becomes diffusive. To confine the electrons in STL for a long time, the typical diffusion length must be comparable to or smaller than the thickness of STL. \citet{Amano2022b} showed that this condition for a particle of velocity $v$ can be written as
\begin{align}
  \frac{1}{6 \eta}
  \left( \frac{m_e}{m_i} \right)
  \left( \frac{D_{\mu\mu}}{\Wce} \right)^{-1}
  \left( \frac{v}{V_{\rm sh} / \cos \tbn} \right)^{2}
  \lesssim 1
\end{align}
with the notations used in this paper. Here $D_{\mu\mu}$ is the pitch-angle scattering rate. The standard quasi-linear theory can be used to estimate $D_{\mu\mu}$ in STL as follows:
\begin{align}
  \frac{D_{\mu\mu}}{\Wce} = \frac{\pi}{4}
  \frac{I(k_{\rm res}) k_{\rm res}}{B_0^2}
  \left( \frac{B_0}{B_{\rm STL}} \right)
  \left( 1 + \frac{\omega_{\rm res}}{k_{\rm res} v} \right),
\end{align}
where the factor $B_0/B_{\rm STL}$ takes into account the magnetic field compression in STL. The wavenumber spectrum is denoted by $I(k_{\rm res})$ with $k_{\rm res}$ being the resonant wavenumber, which is defined by the cyclotron resonance condition $\omega_{\rm res} = -k_{\rm res} v + \Wce$. The last factor $\left( 1 + \omega_{\rm res}/k_{\rm res} v \right)$ in the above equation provides a finite frequency correction to the standard magnetostatic approximation ($\omega_{\rm res}/k_{\rm res}v \ll 1$). Note that, for simplicity, we consider a wave propagating in either parallel or anti-parallel to the ambient magnetic field, and the resonant particle is traveling in the opposite direction of the wave (which gives the negative sign in the resonance condition).

The condition required for the wavenumber spectrum is now readily obtained as:
\begin{align}
  \frac{I (k_{\rm res}) k_{\rm res}}{B_0^2} \gtrsim
  \frac{2}{3 \pi \eta}
  \left( \frac{B_{\rm STL}}{B_0} \right)
  \left( \MAHTF \right)^{-2}
  \left( \frac{v}{\VAe} \right)^2
  \left( 1 + \frac{\omega_{\rm res}}{k_{\rm res} v} \right)^{-1}.
\end{align}
The lower bound then defines the threshold. If we now use the relation $I(k) dk = P(f) df$ between the wavenumber spectrum $I(k)$ and the frequency spectrum $P(f)$, the threshold condition for the frequency spectrum obtained by \citet{Amano2020a} is recovered. Similarly, the threshold for the frequency-integrated wave power used in this paper is calculated as follows:
\begin{align}
  \int_{f_{\rm min}}^{f_{\rm ce}} \frac{P(f)}{B_0^2} df
  & =
  \int_{k_{\rm min}}^{\infty} \frac{I(k_{\rm res}) k_{\rm res}}{B_0^2} \frac{dk_{\rm res}}{k_{\rm res}}
  \nonumber \\
  & \gtrsim
  \frac{2}{3 \pi \eta}
  \left( \frac{B_{\rm STL}}{B_0} \right)
  \left( \MAHTF \right)^{-2}
  \nonumber \\
  & \times
  \int_{k_{\rm min}}^{\infty}
  \left( \frac{v}{\VAe} \right)^2
  \left( 1 + \frac{\omega_{\rm res}}{k_{\rm res} v} \right)^{-1}
  \frac{dk_{\rm res}}{k_{\rm res}}.
\end{align}
The integrand in the last expression may be rewritten by using the resonance condition to yield Eq.~\ref{eq:threshold}. The relation between the resonance velocity and wavenumber $v = (\Wce - \omega_{\rm res})/k_{\rm res}$ indicates that the major contribution to the integral comes from the lower bound $k_{\rm min}$ associated with $f_{\rm min}$.

A few remarks on the limitations of the present analysis are in order. First, we have not taken into account the transformation between the plasma and spacecraft rest frames. Strictly speaking, the frequency appearing in the theoretical expressions should be defined in the plasma rest frame. We have ignored the effect of the Doppler shift because it is typically a small correction as long as high-frequency waves are concerned. Second, we do not distinguish the wave propagation direction. The implicit assumption here is that the wave power is symmetric with respect to the parallel and anti-parallel propagation, and the oblique propagation is ignored. Therefore, the theoretical threshold derived here should only be considered as an order of magnitude estimate.

\bibliography{reference, zenodo}

\end{document}